\documentclass{aa}  

\usepackage{ulem}
\renewcommand{\arraystretch}{1.1}
\usepackage{color}
\usepackage{array}
\usepackage{url}
\usepackage[bottom]{footmisc}
\usepackage{xcolor}
\usepackage{graphicx}
\usepackage{multirow}
\usepackage{booktabs}
\usepackage{caption}
\usepackage{subcaption}
\usepackage{booktabs}
\usepackage{adjustbox}
\usepackage[toc,page]{appendix}
\usepackage{amsmath}
\usepackage{nccmath}
\usepackage{arydshln} 
\bibpunct{(}{)}{;}{a}{}{,}
\usepackage{xcolor}
\usepackage{txfonts}
\usepackage[]{hyperref}
\hypersetup{unicode=true, colorlinks=true, linkcolor=[rgb]{0.53, 0.15, 0.34}, citecolor=[rgb]{0.06, 0.2, 0.65}, filecolor=[rgb]{1.0, 0.13, 0.32}, urlcolor=[rgb]{0.53, 0.15, 0.34}}

\begin{document} 


   \title{The orbital period and inclination of the neutron star X-ray transient MAXI J1807+132}


\author{E. A. Saavedra \inst{1,2} \and
   T. Muñoz-Darias\inst{1,2}      \and
   M. A. P. Torres\inst{1,2}      \and   
   I. V. Yanes-Rizo\inst{1,2}     \and
   M. Armas Padilla\inst{1,2}     \and \\  
   A. Álvarez-Hernández\inst{1,2} \and 
   J. Casares\inst{1,2} \and
   D. Mata Sánchez\inst{1,2} \and   
   S. K. Rout\inst{3,4} \and
   S. Navarro\inst{1,2} 
    }

 \institute{
    Instituto de Astrof\'isica de Canarias (IAC), V\'ia Láctea, La Laguna, E-38205, Santa Cruz de Tenerife, Spain \and 
    Departamento de Astrof\'isica, Universidad de La Laguna, E-38206, Santa Cruz de Tenerife, Spain \and 
    New York University Abu Dhabi, PO Box 129188, Abu Dhabi, UAE \and
    Center for Astrophysics and Space Science (CASS), New York University Abu Dhabi, PO Box 129188, Abu Dhabi, UAE
 }

 
   \abstract
   {
The neutron star X-ray transient MAXI J1807+132 underwent outbursts in 2017, 2019, and 2023. We conducted an $R$-band time series photometry campaign using the \textit{Isaac Newton} Telescope during the 2022 quiescent state. We detected a periodic variation in the light curve that is consistent with ellipsoidal modulation, which allowed us to determine an orbital period of $P_{\rm orb} = 4.258 \pm 0.008$ hr. 
By modelling the light curve, we obtained a binary inclination of $ i = 72\pm5 \, \deg $ and a mass ratio  $q = 0.24^{+0.19}_{-0.14}$ ($68$ per cent confidence level). Furthermore, our analysis indicates the presence of an early M-dwarf companion that contributes between 30 and 50 per cent to the total flux in the $R$ band.
We have extended the previously established absolute magnitude versus orbital period correlation for black hole X-ray transients to neutron star systems. We applied the correlation to MAXI J1807+132, estimating its distance to be $6.3 \pm 0.7$ kpc and its height above the Galactic plane to be $1.6 \pm 0.2$ kpc.
}

   \titlerunning{The ellipsoidal light curve of MAXI J1807+132}
   \authorrunning{Saavedra et al.}
   \maketitle
%
 \section{Introduction}

In low-mass X-ray binaries (LMXBs), a compact object—either a stellar-mass black hole (BH) or a neutron star (NS)—accretes material from a low-mass donor star through Roche lobe overflow. The transferred material forms an accretion disc, where viscous forces govern angular momentum transport \citep{Shakura1973A&A}.  These systems exhibit two distinct accretion regimes: persistent LMXBs, 
distinguished by steady and high X-ray luminosities 
(\(L_X \gtrsim 10^{35}~\mathrm{erg\,s^{-1}}\)), and transient LMXBs (X-ray transients). 
The latter alternate between prolonged quiescent states \citep[\(L_X \lesssim 10^{33}~\mathrm{erg\,s^{-1}}\); see e.g.][]{ArmasPadilla2013MNRAS.428.3083A} and episodic outbursts, during which their luminosity increases to the level of persistent systems.
X-ray transients (XRTs) are typically discovered during outbursts, which are triggered when thermal-viscous instabilities in the disc amplify mass accretion rates to $\dot{M} \sim 10^{-9}$--$10^{-7}~M_\odot~\mathrm{yr^{-1}}$ \citep{Lasota2001NewAR..45..449L}. These events are marked by a sudden rise in luminosity (\( \Delta L \sim 10^{3}\text{--}10^{7} \) erg s$^{-1}$). Outbursts can last from a few weeks to several months \citep{CorralSantana2016A&A...587A..61C, Heinke2024arXiv240718867H}, providing a unique opportunity to study accretion and ejection processes \citep[see e.g.][]{Fender2016LNP...905...65F}. The outburst demise is marked by a drop in the mass accretion rate to \(\lesssim 10^{-11}~M_\odot\,\mathrm{yr^{-1}}\), which causes the system to return to quiescence. In this state, the companion star often accounts for a significant fraction of the infrared and optical emission, although the accretion disc can also make a significant contribution. This is the best scenario for dynamical studies since the fundamental parameters of the binary can be determined, including the compact object mass \citep[see e.g.][for a review]{Casares2017hsn..book.1499C}.

The XRT MAXI J1807+132 (hereafter J1807) was first detected during an outburst in 2017 \citep{Shidatse2017ApJ...850..155S, JimenezIbarra2019MNRAS.484.2078J} and subsequently experienced two additional events, in 2019 \citep{Albayati2021MNRAS.501..261A} and 2023 \citep{Illiano2023ATel16125....1I, Sandeep2025ApJ...978...12R}. NICER observations during the 2019 outburst detected three type I X-ray bursts, confirming the presence of a NS in J1807 \citep{Albayati2021MNRAS.501..261A}. Although these bursts were not of the photospheric radius expansion type, they enabled an upper limit to the distance of 12.4 kpc to be established.

In this paper we present time-resolved $R$-band photometry of J1807 obtained during quiescence over three nights. With these data, we determined the orbital period of the system and modelled the ellipsoidal light curve to infer the binary parameters. We also estimated its distance and elevation above the Galactic plane.

\section{Observations and data reduction} \label{sec:datareduction}
\begin{figure}[h!]
    \centering
    \includegraphics[width=0.99\columnwidth]{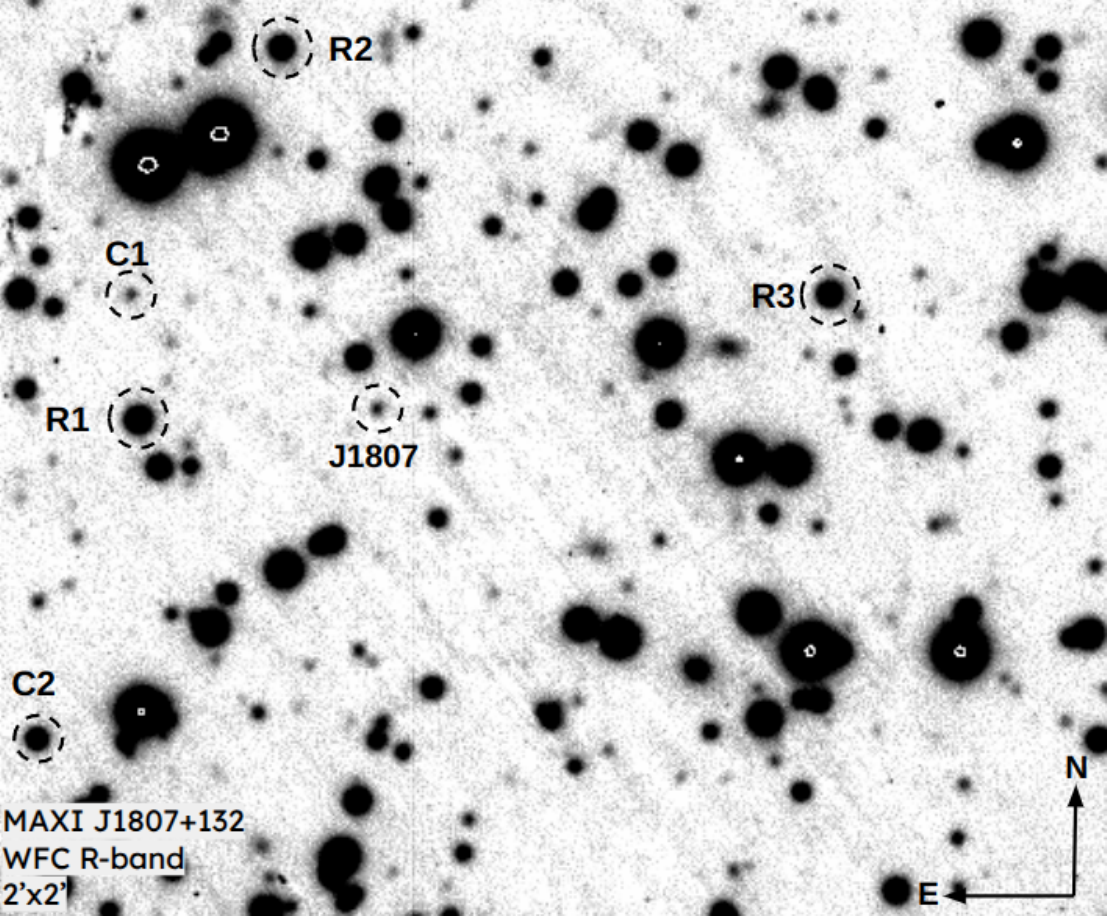}
    \caption{600~s $R$-band image of J1807 in quiescence, obtained with the WFC on the night of 24 June 2022. The target (J1807), the reference star (R), and comparison stars are marked with circles. North is at the top, and east is to the left.}
    \label{fig:ds9}
\end{figure}

We conducted $R$-band photometry of J1807 during quiescence, with the 2.5 m \textit{Isaac Newton} Telescope (INT) at the Roque de los Muchachos Observatory in La Palma, Spain. Observations were carried out on the nights of 24--25 June 2022, and 28 July 2022 (all dates throughout the text are given in UTC). We employed chip 4 of the Wide-Field Camera (WFC), an instrument equipped with four EEV $2048 \times 4096$ pixel CCDs with a plate scale of 0.33 arcsec pix$^{-1}$. We acquired a total of 86 images, each with an exposure time of 600 seconds, using the Harris $R$-band filter. The total observation time was approximately 6.66 hours (40 images), 6 hours (36 images), and 2.33 hours (10 images) on the first, second, and third nights, respectively. The image quality conditions ranged from 1.25 to 1.75 arcseconds on the first two nights, and from 1.5 to 2 arcseconds on the third night. 

The data reduction and photometry were performed with the \texttt{HiPERCAM4} pipeline\footnote{\url{https://cygnus.astro.warwick.ac.uk/phsaap/hipercam/docs/html/}}, with apertures centred on J1807 and five nearby field stars, as shown in \hyperref[fig:ds9]{Fig.~\ref{fig:ds9}}. Owing to the faintness of the source, we applied optimal photometry, which weighted the flux of each pixel according to its expected contribution to the total signal relative to the noise \citep{Naylor1998MNRAS.296..339N}.

\begin{figure}[h!]
    \centering
    \includegraphics[width=1\columnwidth]{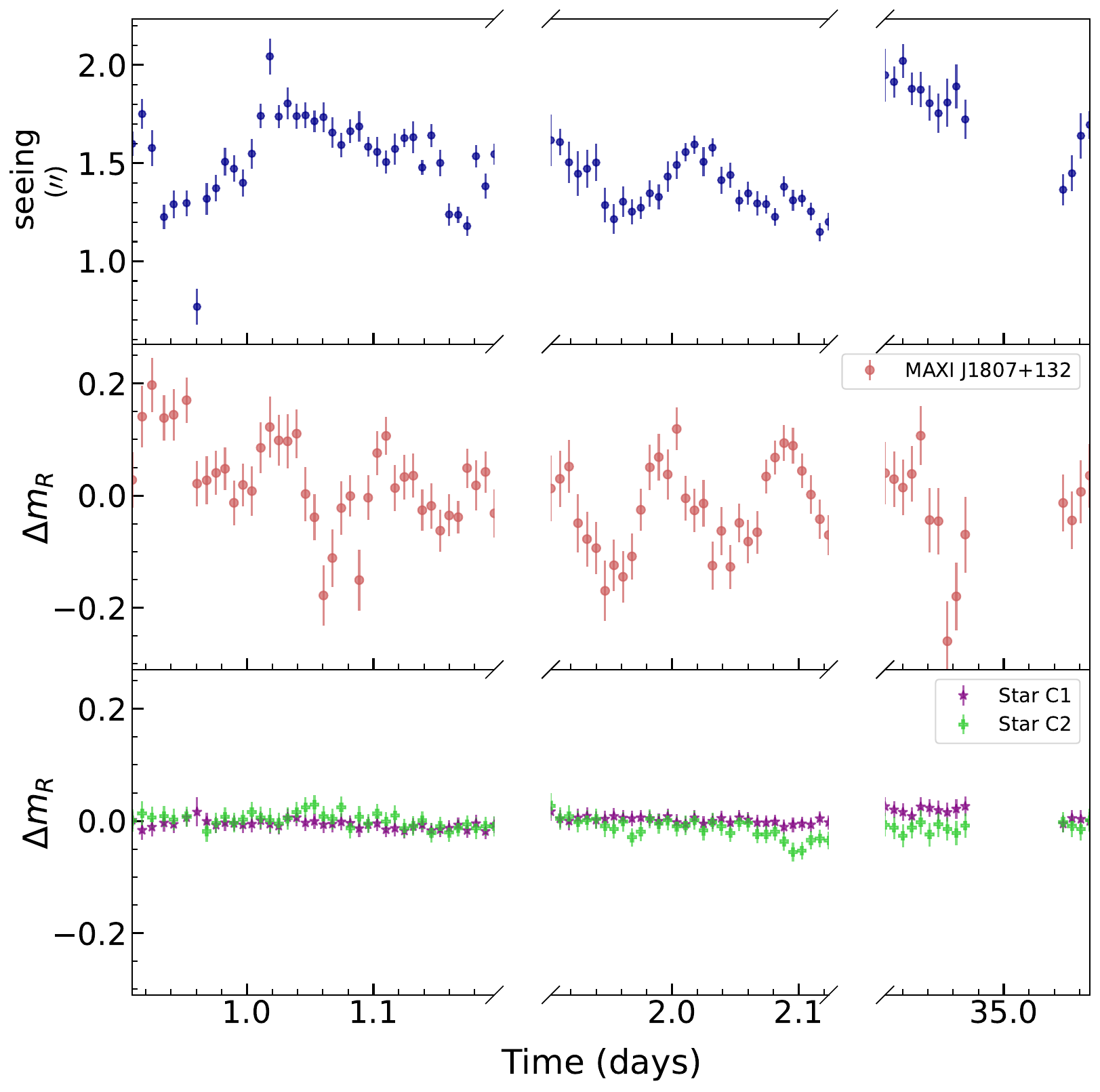}
    \caption{Light curves obtained on 23-24 June 2022 and 27 July 2022. The time is in units of days since HJD 2459754.5.Upper panel: Seeing conditions for each observation. Middle panel: Light curve of J1807. Lower panel: Light curve of the comparison stars marked in Fig.~\ref{fig:ds9}, demonstrating photometric stability.}
    \label{fig:lc_seeing}
\end{figure}

The differential photometry was calibrated using the GSC 2.2 catalogue \citep{GSC2.22001yCat.1271....0S}, with stars labelled R1, R2, and R3 in \hyperref[fig:ds9]{Fig.~\ref{fig:ds9}} as reference stars. We confirmed the stability of these three star by comparative analysis with neighbouring field stars, before extracting the light curve for J1807 and comparison stars (C1 and C2), shown in \hyperref[fig:lc_seeing]{Fig.~\ref{fig:lc_seeing}}. We quantified the variability using the differential magnitude, \(\Delta m_r\), calculated relative to the average quiescent level \(\langle R_{\mathrm{q}} \rangle\) as \(\Delta m_R = R(t) - \langle R_{\mathrm{q}} \rangle\), where \(R(t)\) is the instantaneous \(R\)-band magnitude. The photometric analysis yielded mean $R$ magnitudes and associated RMS variabilities of $21.48 \pm 0.08$ for J1807, $21.18 \pm 0.02$ for star C1, and $19.18 \pm 0.01$ for star C2.

The Pan-STARRS Data Release 2 (DR2) catalogue \citep{Chambers2016arXiv161205560C} provides pre-discovery magnitudes. To estimate the deepest quiescent brightness, we took the faintest minimum values in the $r$ and $i$ bands: $r = 21.65\pm0.20 \, \mathrm{mag}$ and $i = 21.48\pm0.20 \, \mathrm{mag}$. Applying the transformation from Lupton (2005)\footnote{\url{https://classic.sdss.org/dr4/algorithms/sdssUBVRITransform.php}}: $R = r - 0.2936(r - i) - 0.1439$, we obtained $R = 21.46\pm0.15 \, \mathrm{mag}$. This is consistent with our measured value, supporting the conclusion that J1807 was in a deep quiescent state during our observations. 

In order to quantify the possible influence of seeing variations on the photometry, we computed the Pearson linear correlation coefficient, $r$, between the seeing and the differential $R$-band magnitude for each individual night. This yielded $r = 0.04$ ($p = 0.73$), $r = -0.12$ ($p = 0.42$), and $r = -0.07$ ($p = 0.84$) for 24 June, 25 June, and 28 July, respectively. Thus, in all cases we obtained $|r| < 0.2$, with the associated $p$-values far exceeding the conventional significance threshold of 0.05. This indicates that there is no statistically significant correlation between the seeing and the measured flux. We therefore conclude that the observed variability is intrinsic to J1807 and not an artefact due to variable atmospheric conditions.

 \section{Analysis and results} \label{sec:data}

\subsection{Searching for periodicities}

\begin{figure}[h!]
    \centering
    \includegraphics[width=1\columnwidth]{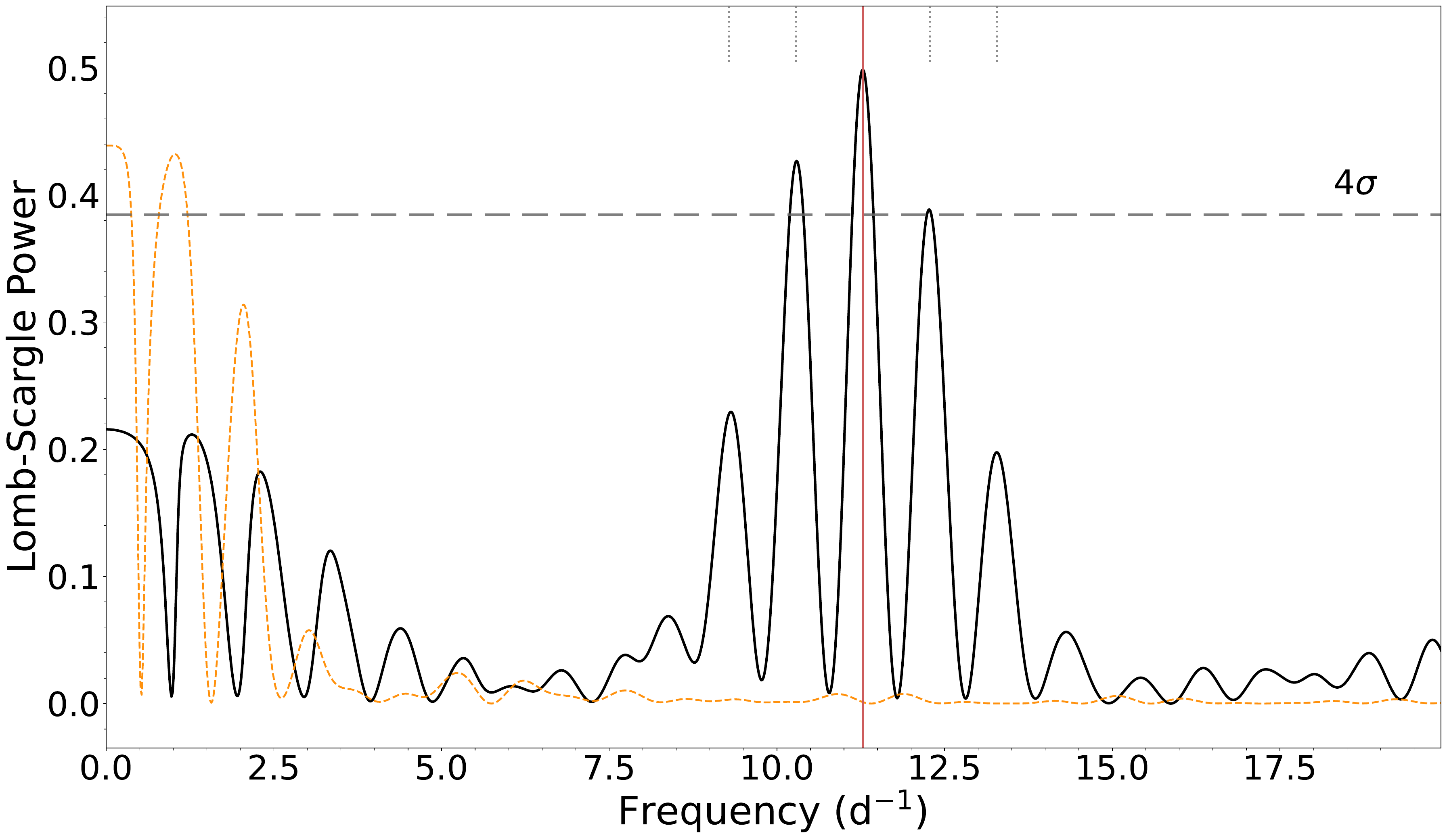}    
    \caption{Lomb-Scargle periodogram of the J1807 light curve for the nights of 24–25 June.  The Lomb-Scargle power of the signal is in black, and the window function in orange. The red vertical line marks the main peak at 2.129 hours; dotted grey ticks at the top indicate the daily aliases ($\pm1$, $\pm2$ d$^{-1}$). The period has a significance greater than $4\sigma$.}
    \label{fig:lombscargle}
\end{figure}

The light curve of J1807 exhibits significant variability, featuring apparently periodic maxima and minima, particularly during the first two nights. In order to minimise one‐day aliasing and avoid the third night, we first computed the Lomb-Scargle periodogram using only the 25–26 June data. In the $0.005-20~\mathrm{d}^{-1}$ range this two‐night analysis reveals a prominent peak at $f_{0} = 11.27~\mathrm{d}^{-1}$ (2.129 h; see Fig.~\ref{fig:lombscargle}).  The sampling window—obtained by applying Lomb-Scargle to a unit‐flux series sampled at the same timestamps—shows the expected daily sidelobes at $f_{0}\pm1,2$ d$^{-1}$; these aliases are indicated by grey dotted ticks at the top of Fig.~\ref{fig:lombscargle} and do not overlap the principal peak.

To assess the significance of the detected frequencies, we computed the false-alarm probability using the \texttt{astropy} routines \citep{Astropy2022ApJ...935..167A}. The three strongest peaks all lie above the $4\sigma$ threshold under the assumption of white Gaussian noise. 
However, this criterion ignores the possibility of a red-noise component. In fact, after subtracting a sinusoidal model at $f_{0}$, we found that the residual power spectrum follows $P(f)\propto f^{\alpha}$ with $\alpha=-0.09\pm0.03$, indicating mild red noise. To quantify the potential contribution of correlated noise, we generated
$10^{5}$ synthetic light curves—each built by adding a sinusoid at $f_{0}$ to $1/f^{\alpha}$ noise produced with the \citet{Timmer1995A&A...300..707T} algorithm—sampled at the same cadence.
We recomputed the Lomb-Scargle periodogram for each realisation and recorded the power at \(f_{0}\); none of the trials reached the observed value, yielding a \(p\text{-value}<10^{-5}\) and confirming that the \(f_{0}\) peak remains significant even in the presence of red noise.

The primary peak was fitted by a Gaussian; therefore, we adopted the mean and standard deviation as the representative value and the $1\sigma$ uncertainty, respectively. We preliminarily determined the period to be $2.129\pm0.004$ hours. Recomputing the periodogram with the third night included produces a consistent Gaussian fit within the uncertainties, confirming the robustness of this period estimate.

\begin{figure}
    \centering
    \includegraphics[width=1\columnwidth]{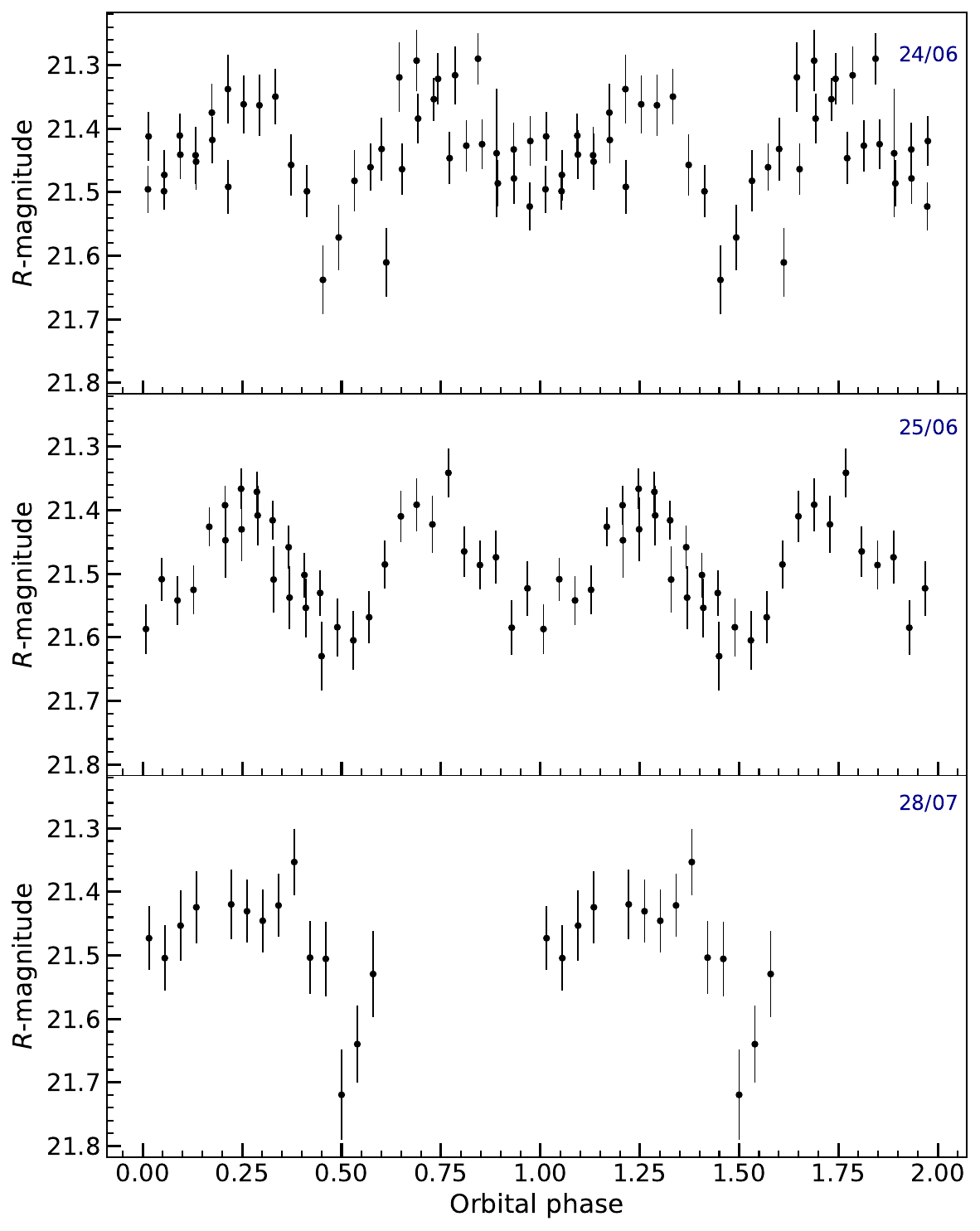}
    \caption{Phase-folded light curves on the 4.258 hrs of J1807. The reference epoch used was $T_{0.5}(HJD)=2459755.572$. The observation dates are given in the upper right of each panel.}
    \label{fig:lcs_fold}
\end{figure}

We independently folded the light curve for each night, testing two periods: the 2.129\,h value derived from Lomb-Scargle analysis, and twice this value, 4.258\,h. Folding with the 2.129\,h period yielded a light curve displaying single-hump modulation, although the minimum depth showed considerable variation from cycle to cycle. Using the 4.258\,h period for folding, however, resulted in a double-humped light curve, which exhibited two minima of unequal depth. This morphology is characteristic of the ellipsoidal modulation expected in quiescence. The gravitational interaction with the NS distorts the companion, and as it orbits, the changing projected area of its ellipsoidal form produces two maxima and two minima per orbital cycle \citep[see e.g.][]{Kopal1954MNRAS.114..101K, Beech1989Ap&SS.152..329B, Wilson1972ApJ...174L..27W, Avni1975ApJ...197..675A}. The deepest minimum is expected to occur at superior conjunction of the companion (orbital phase 0.5) due to the stronger gravitational darkening around the inner Lagrangian point.
This minimum was calculated to occur at $T_{0.5}(HJD)=2459755.572\pm0.002$ (at 1$\sigma$ confidence interval). We then used this epoch to calculate the phases of minima in the other light curves. These were separated by integer multiples of the period and consistently aligned in phase, as shown in \hyperref[fig:lcs_fold]{Fig.~\ref{fig:lcs_fold}}. 

The phase-folded light curves from the first and second nights consistently show two minima within the error margins, with one being deeper than the other. 
We conclude that the period of $4.258 \pm 0.008$\,h represents the orbital period of the system. This determination is based on the consistent features of the light curve, which exhibit characteristic ellipsoidal modulation, including the stable phase alignment of the primary minimum across multiple datasets.

\begin{figure}
    \centering
    \includegraphics[width=1\columnwidth]{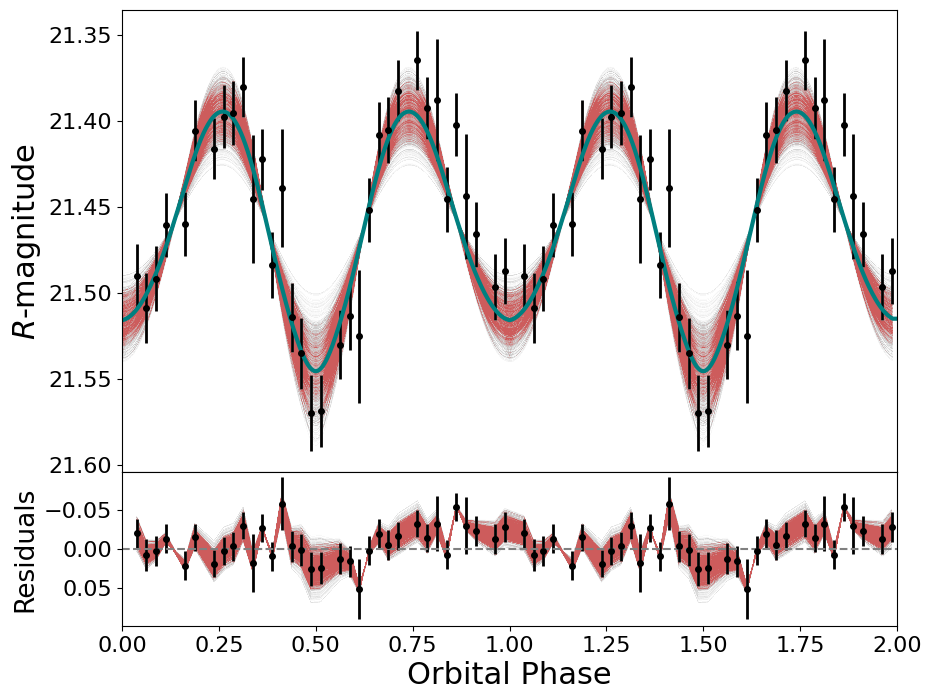}        
    \caption{Light curve modelling. Upper panel: J1807 data (black dots) and the synthetic light curves based on the 68 per cent confidence intervals from the MCMC analysis (red lines). The teal line indicates the best-fit model. Lower panel: Residuals from the best-fit model (black dots) and for each individual synthetic light curve (red lines).}
    \label{fig:lc_fit}
\end{figure}

\subsection{Light curve modelling}
\label{lcmodelling}

Ellipsoidal modulations in quiescent XRTs can be contaminated by aperiodic variability  (flickering) of unknown origin \citep[e.g.][]{Zurita2003ApJ...582..369Z}. To better capture the shape of the modulation in J1807, we grouped the data into orbital phase bins and calculated the weighted median and weighted median error for each bin. Given the variability between nights, using the median allowed us to reduce the impact of outliers. With a phase interval of 0.025 per bin, we found that the characteristic ellipsoidal pattern of the J1807 light curve was accurately represented across the three nights, despite the aperiodic variability present on each individual night.

We used the {\tt XRBinary} code\footnote{Developed by E. L. Robinson, a detailed description of {\tt XRBinary} can be found at: http://www.as.utexas.edu/~elr/Robinson/XRbinary.pdf} to model the light curve \citep[see e.g.,][]{Khargharia2013AJ....145...21K, Gomez2015ApJ...809....9G, Gomez2021MNRAS.502...48G, Ayoze2021MNRAS.507.5805A} and employed the {\tt EMCEE} sampler for fitting \citep{emcee2013PASP..125..306F}. The code computed synthetic light curves of a binary system where a compact primary is accompanied by a co-rotating companion that fills its Roche lobe. The model accounts for the tidal distortion of the companion, which produces ellipsoidal variations, and includes the presence of an optically thick accretion disc that emits as a multi-temperature blackbody. The main parameters that determine the {\tt XRBinary} synthetic light curve are the binary inclination (\(i\)), the orbital period ($P_{\rm orb}$), the mass ratio (\(q\)), the NS mass (\(M_\mathrm{1}\)), the companion's effective temperature (\(T_{\mathrm{2}}\)), and the bolometric luminosity of the disc (\(L_\mathrm{d}\)). The flux spectrum of the donor star was derived from the stellar atmosphere models of \citet{Kurucz1996ASPC..108....2K}, using a non-linear limb darkening formulation from \citet{Claretb2000A&A...363.1081C}, which is valid for surface gravities in the range $0.0 \leq \log g \leq 5.0\,\mathrm{dex}$ and effective temperatures of $3500$--$8000~\mathrm{K}$. The gravitational darkening depends only on the effective temperature of the star, following the formulation in \citet{Clareta2000A&A...359..289C}. 

The model in the {\tt XRBinary} code adopts a cylindrically symmetric disc, with a height profile described by

\begin{equation}
h(r) = H_\mathrm{d} \left( \frac{r - r_{\mathrm{in}}}{R_\mathrm{d} - r_{\mathrm{in}}} \right)^n, \quad r_{\mathrm{in}} \leq r \leq R_\mathrm{d},
\end{equation}

\noindent where \(r_{\mathrm{in}}\) and \(R_\mathrm{d}\) represent the inner and outer radii of the disc, \(H_\mathrm{d}\) denotes the semi-height at the disc's outer edge, and \(n\) is the exponent that governs the height profile. In our model, \(R_\mathrm{d}\) is constrained to be equal to the circularisation radius: 

\begin{equation}
R_\mathrm{d} = (1 + q)(b_1/r)^4, \\
b_\mathrm{1}/r = (1.0015 + q^{0.4056})^{-1},
\end{equation}

\noindent where \(b_\mathrm{1}\) is the distance from the primary star to the inner Lagrangian point, given by \citet{Warner1995cvs..book.....W}. We also evaluated models with the disc reaching the tidal radius \citep{Whitehurst1991MNRAS.249...25W} and found no significant differences in the final results with respect to those using the circularisation radius.

The temperature profile of the disc was set to that of a steady-state, optically thick, viscous disc: 

\begin{equation}
T^4 =  \frac{K}{r^3} \left(1-\left(\frac{r_{\rm in}}{r}\right) ^{1/2} \right) , \quad r_{\mathrm{in}} \leq r \leq R_\mathrm{d}.
\end{equation}

\begin{table}[ht]
\centering
\caption{Fixed and best-fit parameters for the $R$-band light-curve model.}
\resizebox{0.8\columnwidth}{!}{%
\renewcommand{\arraystretch}{1.4} 
\begin{tabular}{ccc}
\hline
\textbf{Parameter} & \textbf{Prior} & \textbf{Best-fit value} \\
\hline
\(T_\mathrm{2}~(\text{K})\)          & Fixed & 3660  \\
\(H_\mathrm{d}\) ($a$)      & Fixed & 0.011 \\
\(r_\mathrm{in}\) ($a$)   & Fixed & 0.02 \\
\(n\)              & Fixed & 1.1 \\
Donor albedo       & Fixed & 0.5 \\
\(i\) (deg)     & [30, 90] & \(72\pm5\) \\
\(\log(L_\mathrm{d})~(\text{erg s}^{-1})\) & [30, 35] & \(32.1\pm0.3\)  \\
\(q\) & [0.02, 0.7]                & \(0.24^{+0.19}_{-0.14}\)  \\
\(M_{\rm NS}~(M\odot)\)        & [0, 3] & \(1.2^{+1.1}_{-0.8}\) \\
\bottomrule
\\\end{tabular}
}
\label{tab:mcmc}
\tablefoot{%
Reported values are medians with $68\%$ confidence intervals. See the main text for details.%
}
\end{table}

\noindent Here, the normalisation constant \(K\) ensures that the temperature distribution matches the bolometric luminosity \(L_\mathrm{d}\) of the disc, and \(r\) is the distance to the primary star. We also investigated a power-law temperature profile (i.e. $T \propto r^{-\beta}$, where $\beta$ = 3/4 corresponds to the asymptotic behaviour of a standard viscous disc) and found consistent results within a $1\sigma$. The disc parameters, such as \(H_\mathrm{d}\), \(r_\mathrm{in}\), and \(n\), were fixed following \citet{vanGrunsven2017MNRAS.472.1907V} and \citet{MataSanchez2021MNRAS.506..581M}. The light curve was symmetric with consistent maxima; therefore, hot spots did not need to be considered, neither on the companion nor in the accretion disc. The full list of model parameters can be found in \hyperref[tab:mcmc]{Table~\ref{tab:mcmc}}. 

We sampled the parameter space within physically reasonable ranges. The orbital period was constrained with a Gaussian prior with mean and standard deviation $4.26 \pm 0.01$ hours, while for the inclination ($i$) we adopted a uniform prior in $\cos(i)$ spanning 30--90 deg (assuming randomly oriented orbits). The disc luminosity ($L_\mathrm{d}$) was sampled using a logarithmically flat prior, with $\log(L_\mathrm{d})$ allowed to vary between 30 and 35. Given that the binary system is composed of a NS and a low-mass companion, we constrained the mass ratio and NS mass using priors of 0.02--0.7 and $<3$ $M\odot$, respectively \citep[see e.g.][]{MuñozDarias2005ApJ...635..502M,Casares2016ApJ...822...99C}.

\begin{figure*}
    \centering
    \includegraphics[width=2\columnwidth]{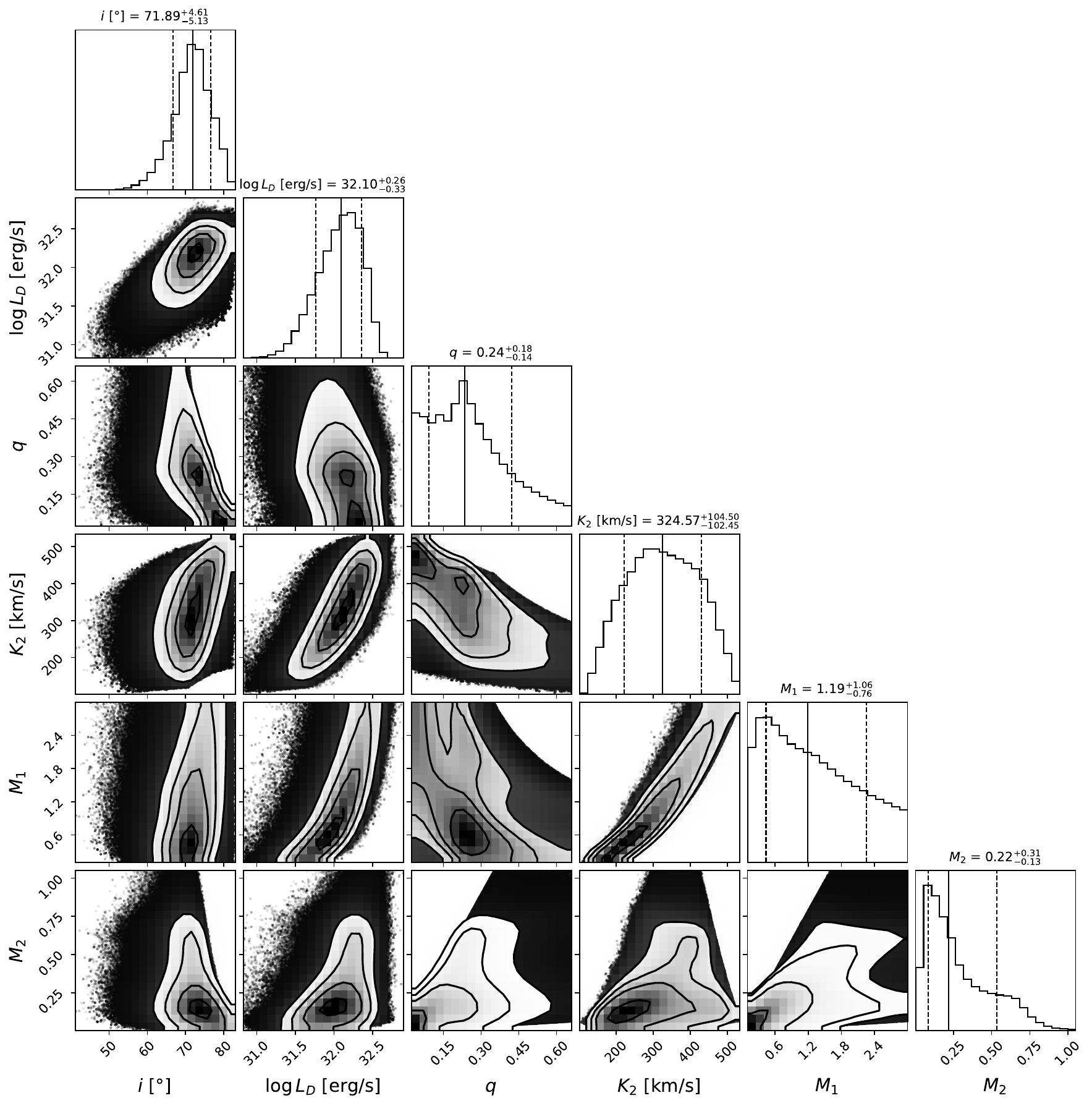}        
    \caption{
Correlation diagrams showing the probability distributions of the best-fit parameters from the MCMC modelling of the ellipsoidal light curve. The 2D plots display contours corresponding to the 68 per cent, 95 per cent, and 99.7 per cent confidence regions. Rightmost panels: Projected 1D distributions of the parameters, with the median indicated by a solid line and the 68 per cent confidence intervals marked by dashed lines. The value of $K_{2}$ is inferred from $q$, $M_{\rm NS}$, $P$, and $i$, whereas the value of $M_{2}$ is inferred from $q$ and $M_{\rm NS}$.}
    \label{fig:mcmcdistribution}
\end{figure*}

We can constrain $T_2$ since Roche-lobe-filling companion stars in LMXBs with orbital periods $\lesssim 7-8$ h are consistent with main-sequence or slightly evolved stars
 (\citealt{Smith1998MNRAS.301..767S}; see also \citealt{Casares2025arXiv250613765C}).
For a Roche-lobe-filling companion star, the orbital period provides an estimate of the mean stellar density via the relation $\langle \rho \rangle = 110 \times P_\mathrm{orb}^{-2}$ g cm$^{-3}$, where $P_\mathrm{orb}$ is expressed in hours \citep{Faulkner1971ApJ...170L..99F}. Our orbital period yields $\langle \rho \rangle = 6.06$ g cm$^{-3}$, consistent with an M1 dwarf \citep{Drilling2000asqu.book..381D}. 
This motivates fixing $T_2$ to the effective temperature of the expected spectral type \citep[$T_{\rm eff}=3660~K$;][]{Pecaut2013ApJS..208....9P}. To evaluate systematic uncertainties, we tested $T_{\rm eff}$ at 3500\,K (the lower bound of our parameter range) and 4000\,K. The orbital inclination varied by only 2--4$^\circ$, demonstrating minimal sensitivity to $T_{\rm eff}$ assumptions.
Additionally, we fixed the donor star's albedo value of 0.5, although varying this parameter has no significant effect on the synthetic light curve.

We ran the Markov chain Monte Carlo (MCMC) sampler for 80000 steps with 136 walkers, discarding the first 20 per cent of the steps as burn-in. An inspection of the MCMC chains confirmed that each parameter converged properly. The resulting model parameters, along with their uncertainties at the 68 per cent confidence interval, are summarised in \hyperref[tab:mcmc]{Table~\ref{tab:mcmc}}. These values were derived from the marginalised posteriors (i.e. the 16th, 50th, and 84th percentiles), and the resulting probability distributions are shown in \hyperref[fig:mcmcdistribution]{Fig.~\ref{fig:mcmcdistribution}}. 
From the posterior distributions of our MCMC analysis, we randomly drew combinations of parameters to generate a set of synthetic light curves. We present these in \hyperref[fig:lc_fit]{Fig.~\ref{fig:lc_fit}} to allow readers to visually assess the quality of the resulting fits. The best-fit model provides a \(\chi^2 = 79.6\) for 67 degrees of freedom (\(\chi^2_{\rm red} = 1.19\)). Our modelling reveals an inclination of $72\pm5$ deg and a bolometric luminosity of the accretion disc of \(\log(L_\mathrm{d})={32.1}\pm0.3\) erg s$^{-1}$. The mass ratio and NS mass distributions yield $q = 0.24^{+0.19}_{-0.14}$ and \( M_{\rm NS} = 1.2^{+1.1}_{-0.8} \, M_{\odot} \) (68 per cent confidence level), respectively.
To determine the companion contribution to the total flux in the $R$ band, we analysed its complete posterior distribution from our synthetic light curves. We constructed histograms from the values obtained in each realisation, extracting the central value and corresponding $1\sigma$ uncertainty. The companion star contributes $42\pm9$ per cent to the total flux.

\section{Discussion}
\label{section:Discussion}

We analysed photometric data of J1807 collected over three nights during its quiescent state. We detected a significant periodicity of $2.129\pm0.004$ hours in the light curves. After folding the data at twice the detected periodicity, we observed two distinct minima, with one being noticeably deeper than the other. Assuming that the deeper minima occurred at orbital phase 0.5 each night, the light curve morphology is consistent with an ellipsoidal modulation. Therefore, we conclude that the orbital period is P$_{\rm orb}=4.258 \pm 0.008$ hours.
Other NS-XRTs with similar orbital periods include XTE J2123$-$058 \citep[$K_2 = 287 \pm 12\,\mathrm{km\,s^{-1}}$, $P_{\mathrm{orb}} = 5.95\,\mathrm{h}$, $i = 69$-$77\,\mathrm{deg}$, $q = 0.49 \pm 0.10$;][]{Casares2002MNRAS.329...29C,Shahbaz2003ApJ...585..443S} and EXO~0748$-$676 \citep[$K_2 \gtrsim 300\,\mathrm{km\,s^{-1}}$, $P_{\mathrm{orb}} = 3.82\,\mathrm{h}$, $i = 75$-$82\,\mathrm{deg}$, $q = 0.11$-$0.28$;][]{MuñozDarias2009MNRAS.394L.136M}.
The phase-resolved $R$-band light curve of J1807 exhibits a peak-to-peak amplitude of $\sim$0.2 mag.  The light curve shows short-timescale stochastic variations ($\sim$0.1 mag amplitude), typically referred to as flickering—a characteristic signature of mass accretion processes. The relatively low amplitude of these variations suggests significant flux contribution from the companion star, which our light curve modelling quantifies at $42\pm9$ per cent of the observed flux. Pre-discovery (i.e. HJD 2455718.4052) Pan-STARRS DR2 photometry \citep{Chambers2016arXiv161205560C} shows significantly larger variability amplitudes of $\sim 0.9$ mag and $\sim 1.7$ mag in the $i$ and $r$ bands, respectively. This variability can be associated with either more intense flickering events or variations in the long-term brightness of the system \citep[see e.g.][]{Cantrell2008ApJ...673L.159C}.

\subsection{Constraints on the binary parameters}

Our light curve modelling analysis (Sect.~\ref{lcmodelling}) constrains the orbital inclination to $72\pm5$ deg. 
The mass ratio and NS mass are loosely constrained. Using the median and 68 per cent confidence intervals from the posterior distribution (Fig.~\ref{fig:mcmcdistribution}), we obtained $q = 0.24^{+0.19}_{-0.14}$ and $M_{\rm NS} = 1.2^{+1.1}_{-0.8} \, M_{\odot}$ ($68$ per cent conf. level).  From these distributions, we derived a companion star mass of $M_2 = qM_{\rm NS} = 0.22^{+0.32}_{-0.13} \, M_{\odot}$. 
This broad $M_2$ range encompasses an M-type companion scenario. Our light curve modelling adopts $T_{\rm eff} = 3660$~K \citep[M1V spectral type,][]{Pecaut2013ApJS..208....9P}. 

We determined the companion-star radial-velocity semi–amplitude (\(K_2\)) using the orbital period together with the posterior distributions of the mass ratio (\(q\)), the NS mass (\(M_{\rm NS}\)), and the inclination (\(i\)) via the mass function

\begin{equation}
f(M) = \frac{K_2^3 \, P_{\rm orb}}{2\pi G} = \frac{M_{\rm NS} \sin^3 i}{(1 + q)^2},
\end{equation}

\noindent finding $K_2 = 325 \pm 105$~km~s$^{-1}$. Similarly, our modelling yields a NS radial velocity semi-amplitude of $K_1 = q\,K_2 = 72^{+41}_{-36}\,\mathrm{km\,s^{-1}}$. Notably, the H$\alpha$ line profile centroid measured across six epochs (five outbursts and one quiescent; see Table 3 in \citealt{JimenezIbarra2019MNRAS.484.2078J}) displays a maximum variation of $63 \pm 35\,\mathrm{km\,s^{-1}}$, consistent with our $K_1$ constraint and suggesting that this disc's emission line broadly tracks the NS motion.  

\subsection{Distance and Galactic elevation}

The quiescent \(r\)-band magnitude and orbital period can constrain the distance to J1807 through the \(M_r-P_{\mathrm{orb}}\) correlation, originally developed for BH-XRTs \citep{Casares2018MNRAS.473.5195C}. As demonstrated in the appendix, this correlation likely applies equally to NS-XRTs. The empirical relation is

\begin{equation}
    M_r = 4.64 \pm 0.10 - (3.69 \pm 0.16) \log P_{\mathrm{orb}} \text{ (d)}.
    \label{eq:correlation}
\end{equation}

\noindent With the faintest Pan-STARRS DR2 pre-outburst magnitude, \(r=21.65\pm0.20\) and \(P_{\mathrm{orb}} = 0.1774 \pm 0.0003\) days, we derived an absolute magnitude of \(M_r = 7.4 \pm 0.2\). The distance to J1807 follows from the distance modulus:

\begin{equation}
    d~(\text{kpc}) = 10^{[0.2 (r - M_r - A_r) - 2]}.
    \label{eq:distance_modulus}
\end{equation}

\noindent Three-dimensional reddening maps, constructed using Pan-STARRS/2MASS photometry and \textit{Gaia} parallaxes \citep{Green2019ApJ...887...93G}, reveal that interstellar extinction towards J1807 increases up to 0.4~kpc, subsequently stabilising at \(E(g-r) = 0.10_{-0.01}^{+0.02}\). This level of reddening corresponds to an \(r\)-band extinction \(A_r = 2.617 \times E(g-r) = 0.26\pm0.02\)~mag, implying a distance of \(6.3 \pm 0.7\)~kpc.

X-ray spectral analysis provides an independent method for estimating distances. 
Analysing NICER spectra from the 2023 outburst, \citet{Sandeep2025ApJ...978...12R} determined an absorption column density of \( N_{\mathrm{H}} = (2.61 \pm 0.02) \times 10^{21}~\mathrm{cm}^{-2} \), considering non-solar abundances. 
Using the relationship \( N_{\mathrm{H}} / E(B-V) = 8.3 \times 10^{21}~\mathrm{cm}^{-2}~\mathrm{mag}^{-1} \) from \citet{Liszt2014ApJ...780...10L}, \( E(B-V) = 0.3 \) mag. 
Assuming a standard Galactic extinction law characterised by \( R_V = 3.1 \) \citep{Cardelli1989ApJ...345..245C, Fitzpatrick1999PASP..111...63F}, we calculated the $r$-band extinction using the relation provided by \citet{Schlafly2011ApJ...737..103S}, yielding \( A_r = 0.71 \pm 0.01 \)~mag.
This estimate is comparable with an alternative value derived using the \( N_{\mathrm{H}} / A_V \) calibration from \citet{Zhu2017MNRAS.471.3494Z}, which gives \( A_r = 0.67 \pm 0.01 \)~mag.
Using these extinction estimates, the derived distances are \( d = 5.06 \pm 0.60 \)~kpc and \( d = 5.17 \pm 0.61 \)~kpc, respectively.

The properties of the companion star offer further constraints. J1807 exhibits a quiescent \(r\)-band magnitude of $21.65\pm0.20$~mag. Accounting for the companion contribution to the total flux (\(42 \pm 9\) per cent), the apparent magnitude of the companion is determined to be \(22.59 \pm 0.32\)~mag. Application of the distance modulus using \(A_r = 0.26\pm0.02\)~mag (from the reddening maps) and adopting absolute magnitudes for companion star of \(M_r* = 8.3\) (M0V) or \(M_r* = 9.0\) (M1V; \citealt{Pecaut2013ApJS..208....9P}\footnote{\url{https://classic.sdss.org/dr4/algorithms/sdssUBVRITransform.php}}) yields distance estimates of 5.5--7.4~kpc and 3.9--5.2~kpc, respectively. These ranges match the distance predicted by the \(M_r\)-\(P_{\mathrm{orb}}\) correlation, reinforcing the classification of the companion as an early M-type dwarf.

The two primary distance determinations (\(6.3 \pm 0.7\)~kpc from reddening maps and \(5.2 \pm 0.4\)~kpc from the 2023 outburst X-ray fitting) differ only $\sim1.7\sigma$. The distance ranges inferred from the companion star properties (3.9--7.4~kpc for M0V--M1V types) overlap with the uncertainty intervals of both primary estimates. In light of this, we decide to adopt as the preferred distance for this study: \(d = 6.3 \pm 0.7\)~kpc. This corresponds to a \(3\sigma\) confidence range of 4.8--7.8~kpc. Using this distance and the Galactic latitude of J1807 (\(b = 15.501\) deg), the Galactic height is \(z = 1.6 \pm 0.2\)~kpc. This places the system significantly above the thin disc (\(z > 1.1\)~kpc).

Using the preferred distance and the 2--10\,keV maximum fluxes observed for the three outburst events \citep{Shidatse2017ApJ...850..155S, Albayati2021MNRAS.501..261A, Heinke2024arXiv240718867H}, we calculated the peak luminosities. Applying a bolometric correction factor of 2.9 \citep{intZand2007A&A...465..953I} and assuming a $1.4\,M_\odot$ NS, we obtain $3 \times 10^{36}$\,erg\,s$^{-1}$ ($\sim$0.017\,$L_{\mathrm{Edd}}$), $2.6 \times 10^{36}$\,erg\,s$^{-1}$ ($\sim$0.015\,$L_{\mathrm{Edd}}$), and $3.7 \times 10^{36}$\,erg\,s$^{-1}$ ($\sim$0.021\,$L_{\mathrm{Edd}}$) for the 2017, 2019, and 2023 outburst, respectively.

\section{Conclusions}

We determined an orbital period of $4.258\pm0.008$ hours for J1807 by analysing photometric data collected over three nights while the system was in quiescence. The detection of ellipsoidal modulation in the light curve allowed us to conduct the first analysis of the binary parameters, establishing an orbital inclination of \(72\pm5\) deg. Additionally, we verified the applicability of the \(M_\mathrm{r} - P_{\mathrm{orb}}\) correlation to NS-XRTs, which enabled us to estimate the distance to J1807 as \(6.3 \pm 0.7\) kpc. This work represents an important step in the characterisation of the dynamical properties of J1807 and lays the groundwork for future spectroscopic studies that will be able to further refine these estimates and obtain the NS mass.

\begin{acknowledgements}
We thank the referee for their thoughtful comments, which substantially enhanced the clarity and robustness of the manuscript.
We acknowledge support by Spanish Agencia estatal de investigación via PID2021-124879NB-I00 and PID2022-143331NB-100.
M.A.P. acknowledges support through the Ramón y Cajal grant RYC2022-035388-I, funded by MCIU/AEI/10.13039/501100011033 and FSE+.
The data were made publically available through the Isaac Newton Group's Wide Field Camera Survey Programme. The Isaac Newton Telescope is operated on the island of La Palma by the Isaac Newton Group in the Spanish Observatorio del Roque de los Muchachos of the Instituto de Astrofísica de Canarias.
The Guide Star Catalogue-II is a joint project of the Space Telescope Science Institute and the Osservatorio Astronomico di Torino. Space Telescope Science Institute is operated by the Association of Universities for Research in Astronomy, for the National Aeronautics and Space Administration under contract NAS5-26555. The participation of the Osservatorio Astronomico di Torino is supported by the Italian Council for Research in Astronomy. Additional support is provided by European Southern Observatory, Space Telescope European Coordinating Facility, the International GEMINI project and the European Space Agency Astrophysics Division.
\end{acknowledgements}

\bibliographystyle{aa}
\bibliography{biblio}

\appendix

\section{\(M_\mathrm{r} - P_{\mathrm{orb}}\) correlation}

A relationship between the absolute $r$-band magnitude (\(M_\mathrm{r}\)) for BH-XRTs in quiescence and their orbital period (\(P_{\mathrm{orb}}\)) was established by \citet{Casares2018MNRAS.473.5195C}. To investigate whether NS-XRTs follow this relationship, we analysed four well-studied systems.

We calculated absolute \(r\)-band magnitudes (\(M_\mathrm{r}\)) through the distance modulus, retrieving the necessary orbital periods and distances for each system. Following \citet{Cardelli1989ApJ...345..245C}, we applied a standard reddening law with \(R_\mathrm{V} = 3.1\) and calculated the extinction in the $r$-band as \(A_\mathrm{r} = 0.748\,A_\mathrm{V}\).

The systems studied were:
\begin{itemize}
    \item Cen~X-4, with \(P=15.0972\pm0.0001\)~h \citep{Torres2002MNRAS.334..233T} and a distance \(d=1.9\pm0.6\)~kpc obtained from the \textit{Gaia} database \citep{Gaia2020yCat.1350....0G}. For this system, we adopted \(E(B-V) = 0.1\) \citep{Blair1984ApJ...278..270B, Shahbaz1993MNRAS.265..655S} and derived \(M_\mathrm{r}\) from the observed magnitude \(r = 17.84 \pm 0.04\) \citep{vandenEijnden2021ATel14317....1V}.

    \item Aql~X-1, having \(P=18.71\pm0.06\)~h \citep{Welsh2000AJ....120..943W} and \(d=6\pm2\)~kpc \citep{Mata2017MNRAS.464L..41M}. In this case, assuming \(E(B-V) = 0.5\) and \((V-R)_0 = 0.8\) (consistent with a K-type companion), we converted \(V\)- and \(I\)-band magnitudes from \citet{Chevalier1999A&A...347L..51C} using SDSS transformation equations\footnote{\url{https://classic.sdss.org/dr4/algorithms/sdssUBVRITransform.php}} to obtain \(r = 20.65 \pm 0.45\).

    \item XTE~J2123-058, with \(P=5.9562\pm0.0001\)~h \citep{Tomsick2002ApJ...581..570T, Casares2002MNRAS.329...29C} and \(d=8.5\pm2.5\)~kpc \citep{Tomsick2004ApJ...610..933T}. For this target, we used \(E(B-V) = 0.13\) \citep{Hynes2001MNRAS.324..180H} with the faintest magnitude found in Pan-STARRS DR2 photometry, giving \(r = 21.92\pm0.15\).

    \item EXO~0748-676, possessing \(P=3.8241\pm0.0001\)~h \citep{Hertz1997ApJ...486.1000H} and \(d=7.7\pm0.9\)~kpc \citep{Wolff2005ApJ...632.1099W}. Here, we used \(E(B-V) = 0.06\) \citep{Parikh2021MNRAS.501.1453P} and \(r=22.12\pm0.04\) \citep{Torres2008ATel.1817....1T}.
\end{itemize}

\begin{figure}[h!]
    \centering
    \includegraphics[width=1\columnwidth]{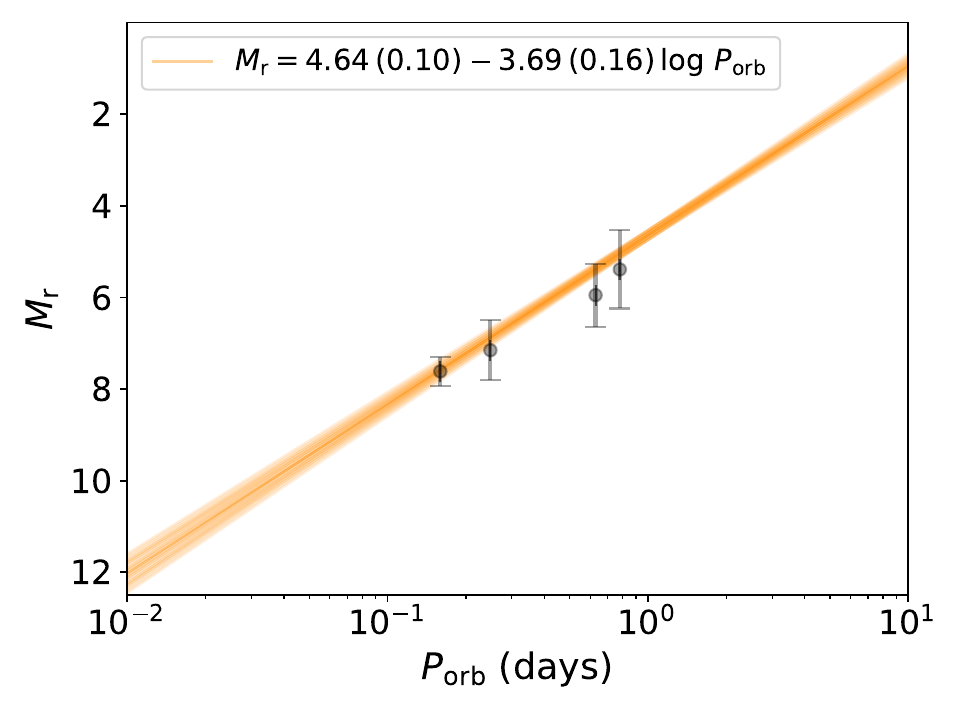}        
    \caption{ \(M_\mathrm{r} - P_{\mathrm{orb}}\) correlation obtained by \citet{Casares2018MNRAS.473.5195C} for BH-XRTs (in orange). The black dots correspond to NS-XRTs, showing that they also satisfy the \(M_\mathrm{r} - P_{\mathrm{orb}}\) correlation.
    }
    \label{fig:correlation}
\end{figure}

Our results, presented in \hyperref[fig:correlation]{Fig.~\ref*{fig:correlation}}, show that the behaviour of these NS-XRTs is consistent with the \(M_\mathrm{r} - P_{\mathrm{orb}}\) correlation established by \citet{Casares2018MNRAS.473.5195C} for BH-XRTs. Therefore, NS-XRTs likely follow a similar trend, at least within the 3 to 16 hour orbital period range tested here.

\end{document}